\documentclass[12pt,paper]{article}

\usepackage{graphics,epsfig}

\newcommand{\degr}{\ensuremath{^\circ}} 

\begin{document}

\pagestyle{empty}  

 \title{\bf Search for cosmic ray sources using muons detected by the MACRO 
experiment}


\maketitle
\begin{center}
{\rm The MACRO Collaboration} \\
\nobreak\bigskip\nobreak
\pretolerance=10000
M.~Ambrosio$^{12}$, 
R.~Antolini$^{7}$, 
G.~Auriemma$^{14,a}$, 
D.~Bakari$^{2,17}$, 
A.~Baldini$^{13}$, 
G.~C.~Barbarino$^{12}$, 
B.~C.~Barish$^{4}$, 
G.~Battistoni$^{6,b}$, 
Y.~Becherini$^{2}$,
R.~Bellotti$^{1}$
% ...
\\
\vspace{1.5 cm}
\footnotesize
1. Dipartimento di Fisica dell'Universit\`a  di Bari and INFN, 70126 Bari, Italy \\
2. Dipartimento di Fisica dell'Universit\`a  di Bologna and INFN, 40126 Bologna, Italy \\
3. Physics Department, Boston University, Boston, MA 02215, USA \\
4. California Institute of Technology, Pasadena, CA 91125, USA \\
5. Department of Physics, Drexel University, Philadelphia, PA 19104, USA \\
6. Laboratori Nazionali di Frascati dell'INFN, 00044 Frascati (Roma), Italy \\
7. Laboratori Nazionali del Gran Sasso dell'INFN, 67010 Assergi (L'Aquila), Italy \\
8. Depts. of Physics and of Astronomy, Indiana University, Bloomington, IN 47405, USA \\
9. Dipartimento di Fisica dell'Universit\`a  dell'Aquila and INFN, 67100 L'Aquila, Italy\\
10. Dipartimento di Fisica dell'Universit\`a  di Lecce and INFN, 73100 Lecce, Italy \\
11. Department of Physics, University of Michigan, Ann Arbor, MI 48109, USA \\
12. Dipartimento di Fisica dell'Universit\`a  di Napoli and INFN, 80125 Napoli, Italy \\
13. Dipartimento di Fisica dell'Universit\`a  di Pisa and INFN, 56010 Pisa, Italy \\
14. Dipartimento di Fisica dell'Universit\`a  di Roma "La Sapienza" and INFN, 00185 Roma, Italy \\
15. Physics Department, Texas A\&M University, College Station, TX 77843, USA \\
16. Dipartimento di Fisica Sperimentale dell'Universit\`a  di Torino and INFN, 10125 Torino, Italy \\
17. L.P.T.P, Faculty of Sciences, University Mohamed I, B.P. 524 Oujda, Morocco \\
18. Dipartimento di Fisica dell'Universit\`a  di Roma Tre and INFN Sezione Roma Tre, 00146 Roma, Italy \\
$a$ Also Universit\`a  della Basilicata, 85100 Potenza, Italy \\
$b$ Also INFN Milano, 20133 Milano, Italy \\
$c$ Also Istituto IASF/CNR, 40129 Bologna, Italy \\
$d$ Also Universit\`a  di Trieste and INFN, 34100 Trieste, Italy \\
$e$ Also Dipartimento di Energetica, Universit\`a  di Roma, 00185 Roma, Italy \\
$g$ Also Institute for Space Sciences, 76900 Bucharest, Romania \\
$h$ Macalester College, Dept. of Physics and Astr., St. Paul, MN 55105 \\
$i$ Also Department of Physics, Pennsylvania State University, University Park, PA 16801, USA \\
$l $Also Dipartimento di Fisica dell'Universit\`a  della Calabria, Rende (Cosenza), Italy \\
$o$ Also Dipartimento di Scienze e Tecnologie Avanzate, Universit\`a  del Piemonte Orientale, Alessandria, Italy \\
$p$ Also University of Minnesota Duluth Physics Department, Duluth, MN 55812 \\
$q$ Also Dept. of Physics, MIT, Cambridge, MA 02139 \\
$r$ Also Intervideo Inc., Torrance CA 90505 USA \\
$s$ Also Resonance Photonics, Markham, Ontario, Canada\\
\end{center}

\newpage
\begin{abstract}

  The MACRO underground detector at Gran Sasso Laboratory recorded 60
  million secondary cosmic ray muons from February 1989 until December 
2000.
  Different techniques were used to analyze this sample in search for
  density excesses from astrophysical point-like sources. No evidence
  for DC  excesses for any source in an all-sky survey is reported. In
  addition,  searches for muon excess correlated with the known binary
  periods of Cygnus X-3 and Hercules X-1, and searches for
  statistically  significant bursting episodes from known gamma ray
  sources are also proved negative.

\end{abstract}

\section{Introduction} 

The interest in the search for cosmic ray point sources  identified
from a  measurable flux of underground muons has historical
motivations  mainly because of the ``CygX-3 saga''. 
Cygnus X-3 (CygX-3) is a galactic binary system well studied in all
types of electromagnetic radiation, most notably in the X-rays.
At  $\gamma$-ray energies, $E\geq 10^{15}$~eV, the Kiel
Extensive Air  Shower array (EAS) initially reported an excess of
events from CygX-3 correlated  with its 4.8 hour binary
period\cite{Samorski:zm}.
 This observation appeared to confirm
previous results at lower energies \cite{Crimea,JPL}. Subsequently,
several  groups operating EAS array experiments and atmospheric
\v{C}erenkov  telescopes confirmed the signals with 
  different  statistical significance and at different energy
thresholds  \cite{Lloyd-Evans,Kifune,Baksan1,Cygnus1,Akeno2,Muraki}.
These results prompted a new generation of more sensitive experiments
using new techniques. The new experiments reported  clear evidence of TeV
$\gamma$-ray emissions from many galactic ($i.e.$, Crab and SN1006
\cite{Cangaroo,Whipple,Hegra}) and the extragalactic AGN-like
($i.e.$, Mrk421 and Mrk501 \cite{mkn421a,mkn421b}) sources.
Finally, in 1990-1991 the new generation detectors
CASA-MIA\cite{Borione:1996jw} and CYGNUS\cite{Alexandreas:1992ek}
put stringent upper limits to the $\gamma$-ray signal from CygX-3,
which excluded early observations. 

Soon after the initial EAS detection from CygX-3, two
underground  experiments (Soudan 1 \cite{Soudan1} and NUSEX
\cite{NUSEX})  reported excesses in TeV muons pointing back to the
CygX-3 direction when correlating event arrival time with the known
orbital period of the source. These detections suggested new physics
beyond the Standard Model since muons are the products of the decay of
charged pions and kaons, which are produced by primary Cosmic Ray (CR)
interactions  with atmospheric nuclei. Charged CR nuclei cannot
propagate directly  from CygX-3 across 10~kpc to the Earth   and there
is only a very low probability of UHE $\gamma$-rays producing TeV
   muons underground\cite{Halzen:1996ik,Bhattacharyya:1995cv}.
Investigations by other underground detectors
\cite{kamiok,frejus,imb,homestake},  however failed to detect any
 significant muon excess from that source or any other.

The MACRO experiment, which ended its operations life in
 December 2000,  was a large area underground detector able
 to reconstruct muon arrival directions to very high
 accuracy \cite{MACROMOON}.  The apparatus started operations in
 1989 and was completed in April 1994.  A preliminary
 investigation  using a limited data sample of $\sim 1.8 \times 10{^6}$
 muons collected by its first and second supermodules was
  published in 1993 \cite{MUASTRO_PAP}.
Since then, MACRO has increased the muon sample by a factor of 30,
its pointing capability has been accurately determined and its
direction  reconstruction capability carefully studied.

In this paper we present the final results on the search for a muon
excess with respect to the evaluated background by surveying the sky
in  declination from -15\degr~ to 90\degr~ and from several candidate
UHE gamma  ray emitters. We also present  a more sensitive search
for  CygX-3 and HerX-1 using their known periodicities. Finally we
searched  for bursting behavior of Mrk421 and Mrk501.

\section{The MACRO detector}

MACRO was a multipurpose underground detector located in the Gran
Sasso  underground Laboratory (LNGS-INFN). It was designed to search
for  rare events in the cosmic radiation and its sensitivity was
optimized  to detect
supermassive magnetic monopoles. The detector \cite{MACRO_OLD,MACROdet} had a
modular  structure and dimensions of $76.5 \times
12 \times 9.3 $ m$^{3}$ with a total acceptance for an isotropic flux
of about 10$^4$ m$^2$sr.  The rock overburden had a minimal depth of
3150 m.w.e and an average of 3700 m.w.e. 

The minimal energy of surface muons that trigger the MACRO apparatus was
about 1.3 TeV. The rock coverage was very irregular and the actual
slant  depth under which a source is observable was taken into account
when  evaluating the muon flux at the surface.
The full detector acceptance for downgoing muons with zenith angles
$\leq 72^\circ$ was about 3100~m$^{2}$sr. 

The detector has worked with
different  configurations starting with 1 supermodule at the 
beginning  and finally with the full configuration consisting of 6
supermodules and the upper half called the ``Attico''\cite{MACROdet}.
The active detection elements are planes of streamer tubes for
tracking  and liquid scintillation counters for fast timing. The lower
half of the detector is filled with streamer tube planes alternating
with  trays of crushed rock.
The upper part is hollow and contains the electronics racks and work
areas.   There are 10 horizontal streamer tube planes in the bottom
half of  the detector, and 4 planes on the top, all with wires and
27$^\circ$  strips, providing stereo readout of the detector hits. Six
vertical planes of streamer tubes cover each side of the detector.

\section{Data selection} 

The data used in this analysis were collected in the period May 1989-
December 2000. The total number of recorded muons is approximately 60
million.

In order to optimize the quality of the tracking reconstruction and to
eliminate periods when the detector was malfunctioning, we applied
several  cuts on a run-by-run basis.
These run cuts include a check on the efficiency of the streamer tubes
system whose data were used for the track reconstruction, and a check
on  the event rate. The  streamer tube efficiency  was obtained using
the sub-sample of tracks crossing all the 10 lower streamer tube
planes;  runs having an average efficiency smaller than 90$\%$ for the
wire view and 85$\%$ for the strip view were discarded. 
As the detector configuration changed during the data taking, the
average  value of the counting rate was computed for each run. Those
with  abnormal rates, $i.e.$, runs having muon counting rate that
deviated by  more than $\pm 3\sigma$ from the average, were cut.

In addition to the described run cuts, we applied event cuts by excluding:

\begin{itemize}

\item  events having zenith angles larger than 72$^{\circ}$; the cut
  is due to  the large uncertainties in the rock depth crossed by these events;

\item events whose reconstruction in one of the two streamer tube
    stereo views  is missing; 

\item  muons in a bundle  with multiplicity larger than 2; this cut
  is necessary  because  the reconstructed direction in space for
  high multiplicity  events is not completely reliable;

\item events that had no arrival time given by the Universal Time Clock (UTC); 

\item  events that do not cross at least 3 planes of the lower part of
  the  apparatus; this cut excludes low energy muons \cite{backsca}
coming from possible secondary  interactions and it improves the angular 
resolution.
The minimal muon energy to satisfy this cut is 1 GeV. 

\end{itemize}
These cuts reduce the sample to 49.9 million well-reconstructed
single  and double muon events during 74,073 hours of livetime. Table
\ref{tab:sample}  shows, for different data taking periods, the
detector  configuration, the number of events which survived the cuts
and the effective livetime.

\section{Background estimation} 

Several methods were suggested to evaluate the background when looking
for an excess of counts from a fixed direction in the sky
\cite{Alexandreas:1992ek,bo1a,bo1b}. 
Many experiments simply consider, as an estimation of the background,
the average value of the counts of the surrounding sky bins with
respect  to a selected one \cite{bo2}.  
An alternative way is to average the counts of all the bins at the
same  declination $\delta$, except the chosen one in a particular
value of  the right ascension (RA).  The assumption that cosmic rays
arrive uniformly from any direction in the sky is implicit in this method.
For an underground detector both methods require that 
 the unequal distribution of the overburden and the
dead-time of the apparatus be taken carefully into account \cite{kamiok}.

\begin{figure}[p]

\begin{center}
\resizebox{15cm}{15cm}{\includegraphics{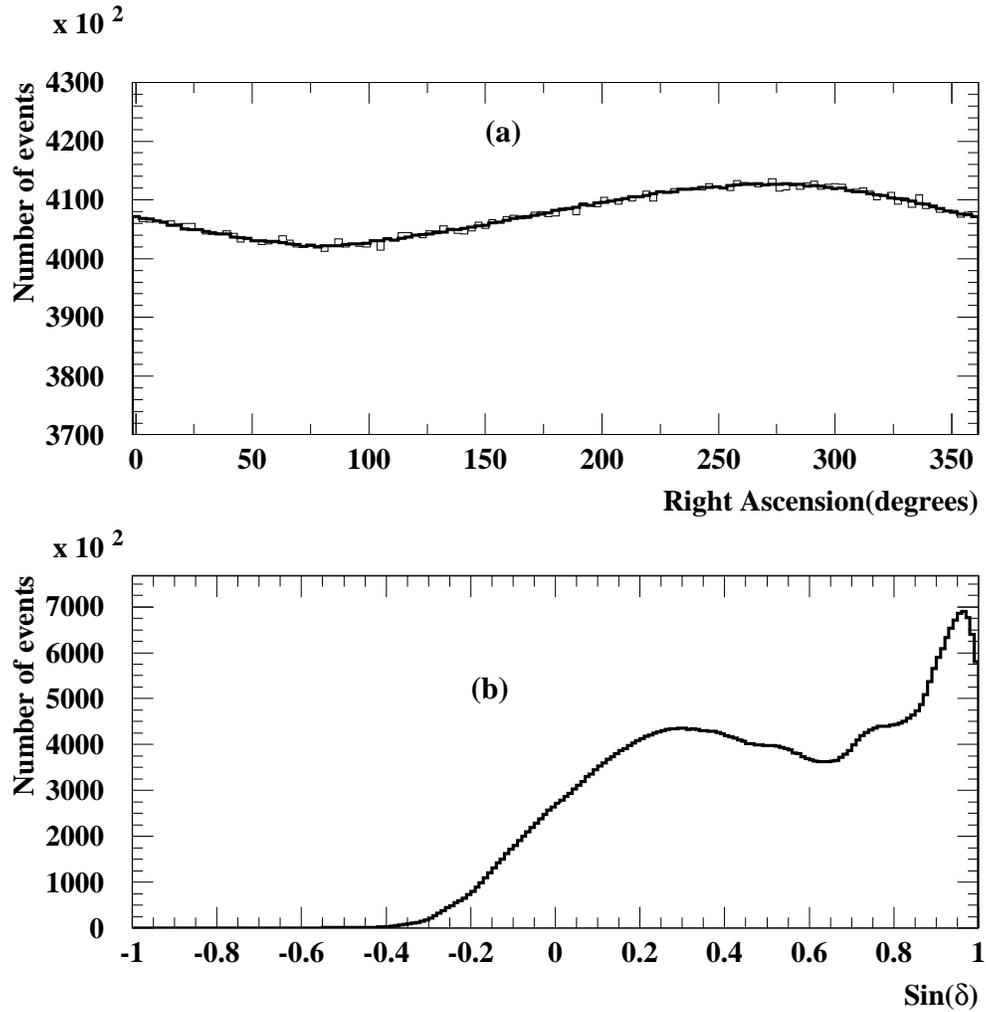}}

\end{center}
\caption{
(a)Right Ascension distribution of the sample of 49.9
  million muons. (b) Distribution of the $\sin\delta$ for the same
  sample of events. The normalized simulated background is
   superimposed in both
  figures, however in (b) they are too close to distinguish. 
}
\label{fig:radec}
\end{figure}

In our analysis, we adopted a different approach.
Assuming that the arrival directions and the arrival time of
underground  muons are uncorrelated variables, we 
  constructed sets  of simulated events by randomly coupling the
times and  the directions of each event in local coordinates.
A mandatory requirement for this method is a good accuracy in the 
  measurement of  the arrival direction and of the recorded time.
For this reason, we excluded from the analysis all the runs and
the events with  errors in the readout time,  as described in
  the previous  section.
We determined that the optimum  number of background events to be associated
with each real event, minimizing the computer processing time, is 25. A
total number of $1.3\times 10^8$ simulated muons were generated as background. 
Figure \ref{fig:radec}  presents the distribution of observed
events  $vs.$ the right ascension RA (a) and the sine of the
declination  sin$\delta$ (b) together with the simulated background.
The fluctuations in the simulated distributions are small compared
with  those in the data.
The unevenness in RA is due to the effect of the dead times.
 The sin$\delta$ distribution reflects the shape of the mountain and
 also the exposure of MACRO (see \cite{MUASTRO_PAP}).

\section{Best sky bin definition}

It is usual for astronomical telescopes to  define a point spread
function  (PSF) that in most cases is a simple bi-dimensional normal
distribution. The PSF can be used to define the
optimal  source bin, $i.e.$, the half-angle $\theta$ of a cone
centered on  a source giving the maximum signal over background ratio
$\frac{S}{\sqrt{B}}$.
For a normal PSF
distribution 
with
variance $\sigma$ it is shown in \cite{Alexandreas:1992ek} that  the
best  value of the half-angle of a cone 
is  $\theta=1.58 \sigma$. This cone contains, on average, 72$\%$ of
the total  number of events.

The MACRO PSF was defined using the double muon sample.
A double muon event (as in general multiple muon events) is produced by  an 
interaction of a primary nucleus at the top of the atmosphere; muons are 
expected to arrive at the earth surface practically parallel.  When 
reaching an 
underground detector, the reconstructed spatial directions of these two muons 
differ as a consequence of the independent scattering of each muon.
We can therefore estimate the MACRO PSF by using as a variable the
distribution of the angular separation of the reconstructed muon
directions  $\theta$ for 
double muon events. This value must be divided by a $\sqrt{2}$ factor to 
account for the two independent scatterings 
\cite{MUASTRO_PAP}. 

\begin{figure}[p]

\vspace{1cm}
\resizebox{15cm}{15cm}{\includegraphics{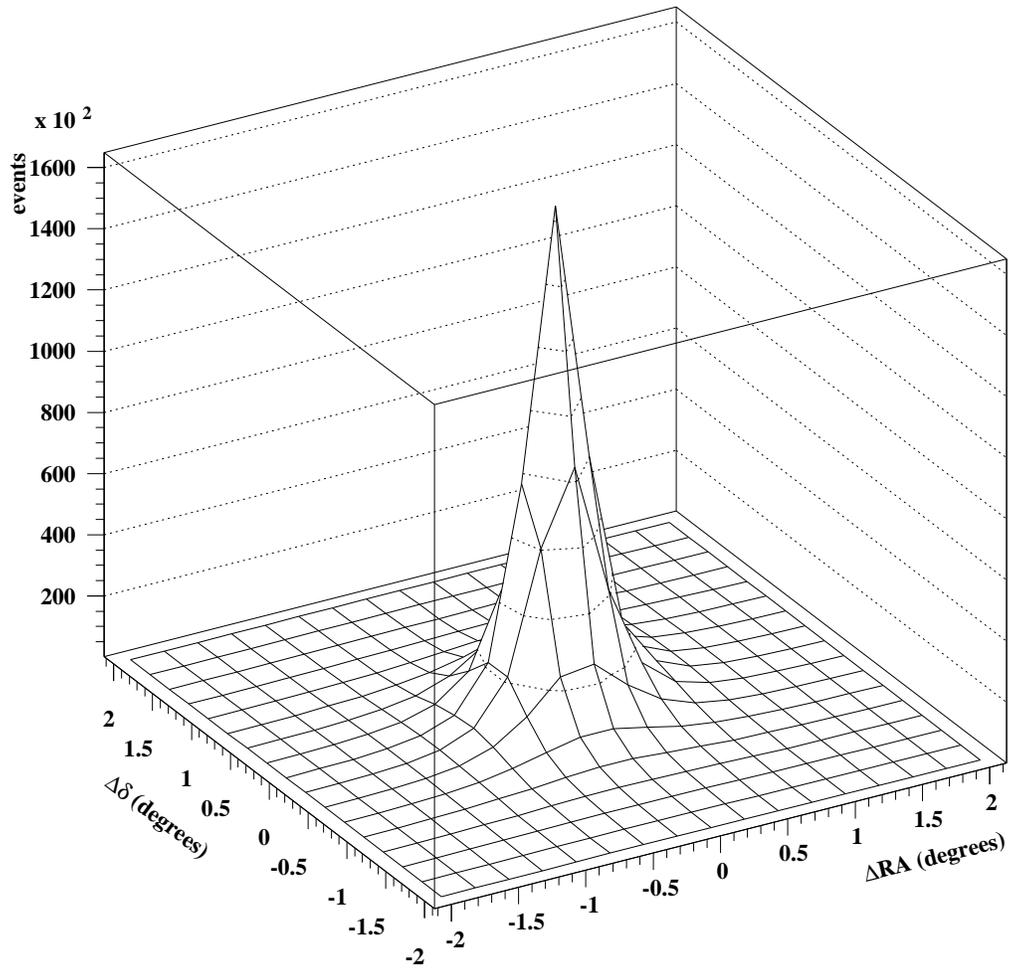}}
\caption{
MACRO Point Spread Function, derived from  the measured
differences in right ascension and declination coordinates of each muon
in double muon events divided by $\sqrt{2}$.
}
\label{fig:PSF}
\end{figure}

We found \cite{MUASTRO_PAP, MACRO_OLD,MACROdet} that 50$\%$ of double muons 
events  are contained in a cone of 0.5\degr ~half-angle,  and that
72$\%$  of events are 
contained  in a cone of 1.05\degr ~half-angle. Initially, 
this result suggests that the best value for the MACRO search bin  (assuming 
that the $\theta$ is normally distributed) is an half-angle cone of 
1.05\degr/$\sqrt{2}$. 
To produce the bidimensional ($RA,\delta$)  distribution representing our PSF 
(shown in Fig.~\ref{fig:PSF}) we used the differences between the reconstructed 
muon directions in RA and declination $\delta$, divided by $\sqrt{2}$.
The main contribution to the differences between the reconstructed  directions is 
the multiple Coulomb scattering (MCS) of muons in the rock. The PSF shown in 
Fig.~\ref{fig:PSF} has a non-gaussian shape, with more events for small 
displacements and  longer tails. The latter were produced by stochastic 
interaction process in the muon energy loss. 

\begin{figure}[p]

\vspace{1cm}
\resizebox{15cm}{15cm}{\includegraphics{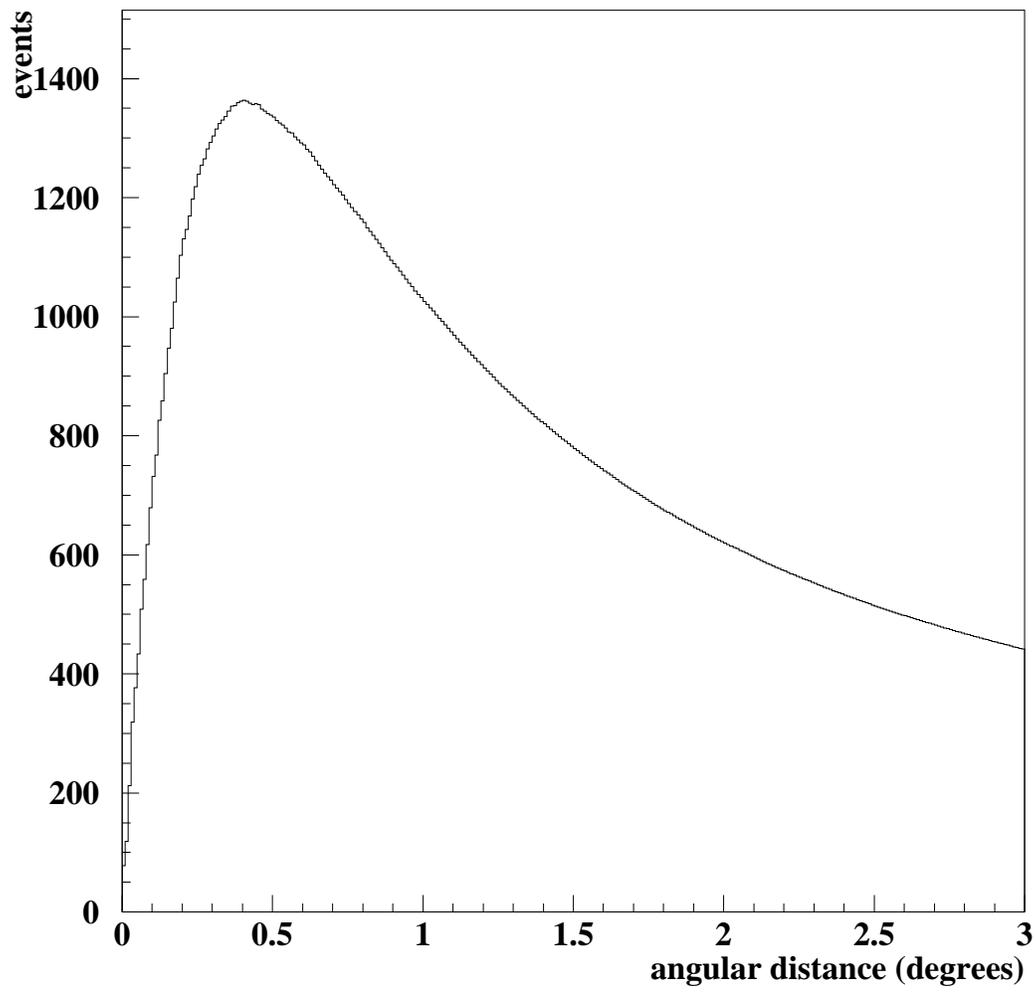}}
\caption{Simulated  $\frac{S}{\sqrt{B}}$ function for the MACRO PSF 
vs the space angle from the source center.
 The maximum value occurs about at $0.45^\circ$.}
\label{fig:S_N_PSF}
\end{figure}

To obtain the value of the best search cone angle, we performed a dedicated 
Monte Carlo.  A source was simulated in a particular ($RA,\delta$) cell, and
the events were extracted according to the PSF of Fig.~\ref{fig:PSF}.
The background events were generated according to a flat density
distribution.  We calculated the ratio $\frac{S}{\sqrt{B}}$ as a
function  of the search cone angle, as shown in
Fig.~\ref{fig:S_N_PSF}.  The maximum of the $\frac{S}{\sqrt{B}}$ ratio
is reached for a search cone angle of $\sim$ 0.4-0.5\degr,  which thus
will represent the best choice value for MACRO.

As an independent check of the validity of the simulation, we studied
the signal due to the Moon shadowing effect (in this case, a lack of
events)  measured by our detector \cite{MACROMOON}. The cumulative
number  of missing events obscured by the Moon's disk, as function of
the  angular distance from the Moon center, shows 
maximum evidence for a deficit using 
a cone of 0.45\degr ~half-angle.

From all these indications the best value for the search cone angle was chosen 
to 0.5\degr.  
However, in the analyses with very low bin contents such  as those
used in the search for 
flaring activities, we use an enlarged search window due to poor 
statistics.

\section{Search for DC sources}

In the search for a steady excess of muons from any direction of the sky (DC 
point-like sources search) we performed an all-sky survey without $\emph{a 
priori}$ assumptions.
We divided the sky into 37176 bins of equal solid angle ($\Delta$$\Omega$= 
2.3$\times$ $10^{-4}$\,sr; $\Delta$RA = $1^{\circ}$, $\Delta$sin$\delta$ = 0.013). 
These bins have the same $\Delta$$\Omega$ as a narrow cone of half-angle  
$0.5^{\circ}$.
We examined the sky bins looking for significant deviations from the
simulated background. The deviation was defined as 
$\sigma= {(n_{obs}-n_{exp})\over{\sqrt{n_{exp}}}}$, where $n_{obs}$ is
the number of events observed in each bin and $n_{exp}$ is the expected
number  of background events in that bin.

\begin{figure}
\vspace{1cm}
\resizebox{15cm}{15cm}{\includegraphics{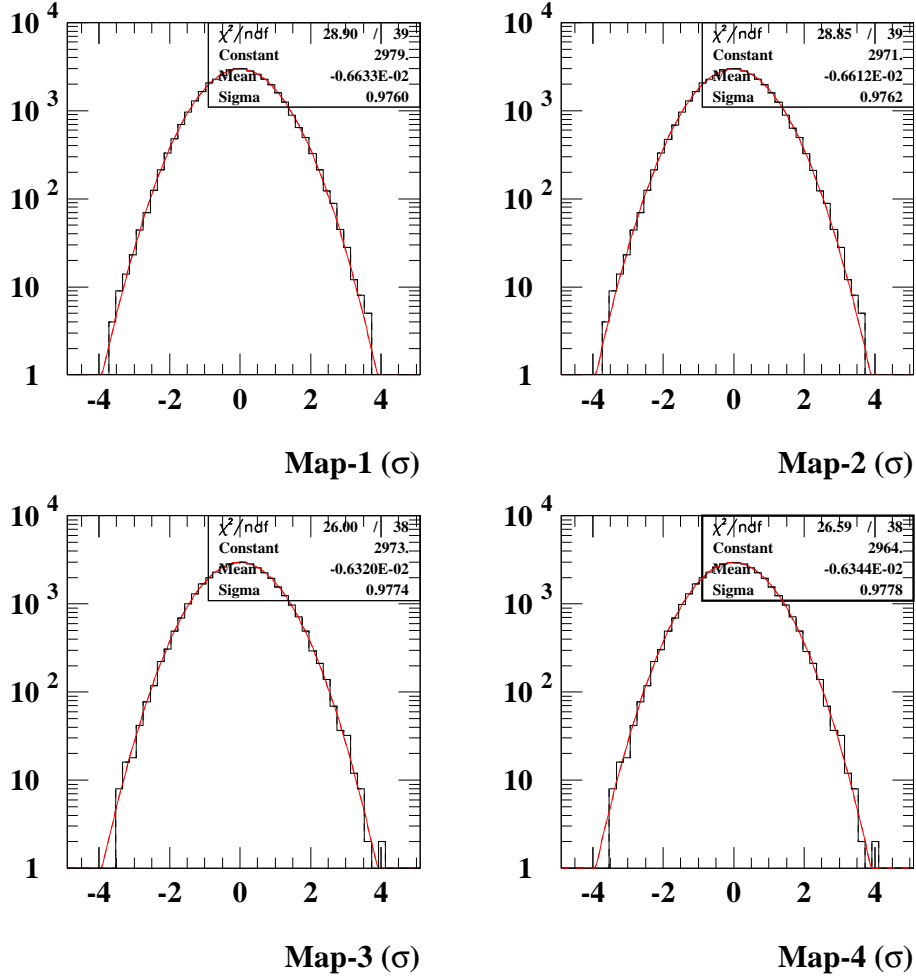}}

\caption{
Map-1 shows distributions of the quantity $\sigma=
{(n_{obs}-n_{exp})\over{\sqrt{n_{exp}}}}$,  where $n_{obs}$ is the
number of events  observed in each sky bin and $n_{exp}$ the expected
number of background events in that bin. Each bin has the same solid
angle  $\Delta\Omega=2.3\times10^{-4}\,{\rm sr}$. 
Map-2 through Map-4 were obtained by the same procedure of Map-1,
but with shifts in RA 
by +0.5$^\circ$(Map-2), or in $\sin\delta$ by   +0.013 (Map-3) or both
(Map-4). 
}
\label{fig:sky}
\end{figure}

\begin{figure}
\vspace{1cm}
\resizebox{16cm}{12cm}{\includegraphics{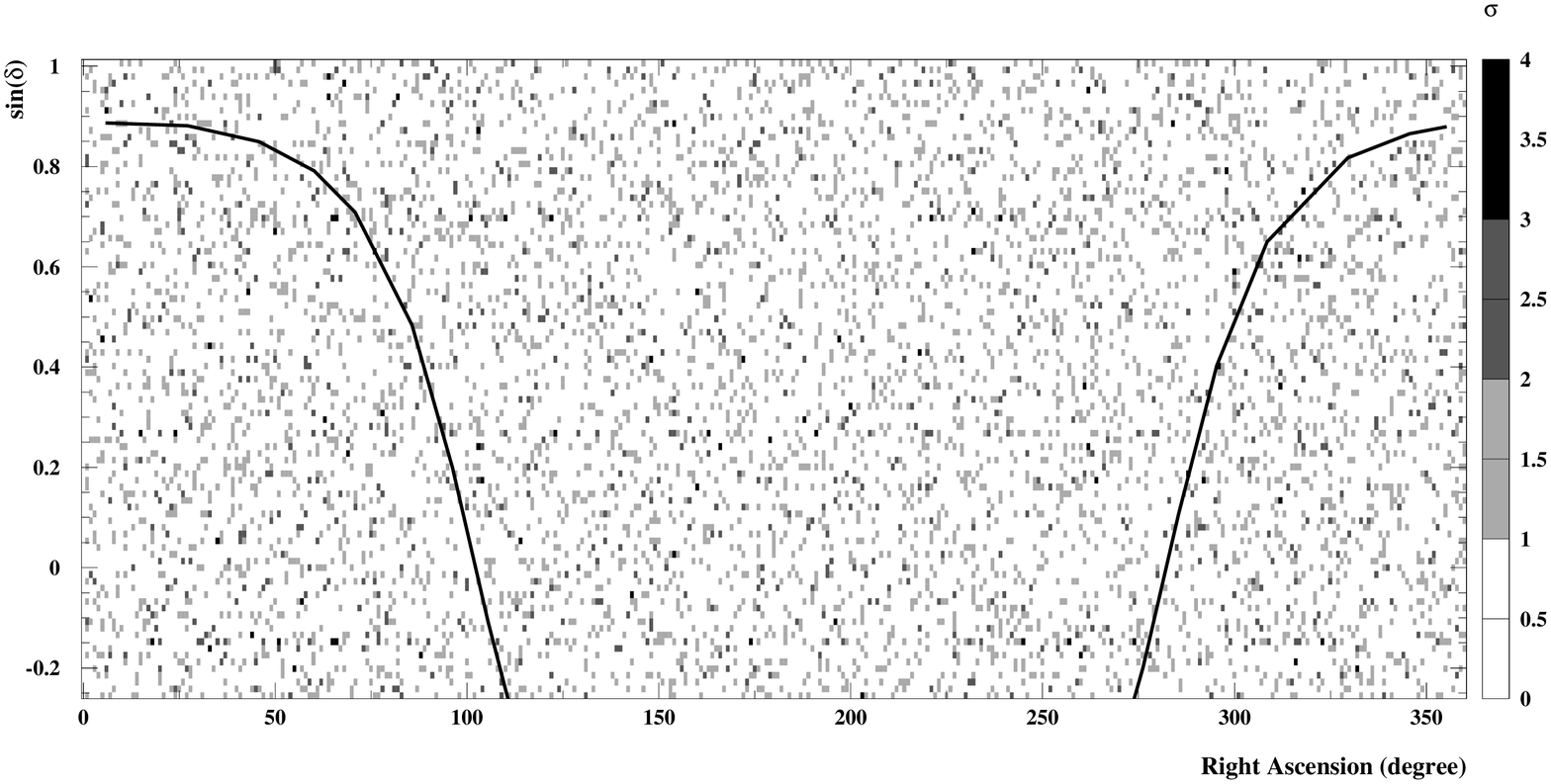}}

\caption{
Bidimensional distribution (in right ascension and sine of the
declination) of the positive  value of 
$\sigma={(n_{obs}-n_{exp})\over{\sqrt{n_{exp}}}}$,  where $n_{obs}$ is  the
number of events  observed in each bin and $n_{exp}$ the expected
number of background events  in that bin.  The line superimposed indicates the 
galactic plane. No point sources are seen.
}
\label{fig:bidev}
\end{figure}

We used three different sky grids, each displaced by half bin
width  in RA, in sin($\delta$) and in both coordinates, to take into
account  the possibility that a source can be located at the edge of
one of  the bins.
Fig.~\ref{fig:sky} shows the distributions of the deviations for the
four  sky-maps with the best-fit Gaussian function superposed. 
The  positive deviations in the first map are reported in the
bidimensional map  in RA and sin$\delta$ of Fig.~\ref{fig:bidev}.
 No
particular  pattern or clustering of positive deviations is observed.
The line indicating the galactic plane is also shown. We
conclude that there is no  evidence for a steady source  emitting
muons  in our data.

We calculated the 95$\%$ C.L muon flux for all sky bins using
the  formula \cite{MUASTRO_PAP}:
\begin{equation}
\mathop{\mathrm J^{stdy}_{\mu}(95\%)}\leq
\frac{\displaystyle n_{\mu}(95\%)}{\displaystyle KA_{eff}T_{exp}}
\mathop{\mathrm{cm^{-2}s^{-1}}}\\ \\
\label{eq:helene}
\end{equation}

\noindent where: 
\begin{enumerate}
\item[-] $n_{\mu}$(95\%) is the upper limit for the number of muons in
  the  bin at 95\%C.L., where $n_{obs}$ and $n_{exp}$ are respectively
  the  number of  observed and expected events in that
  bin. $n_{\mu}$(95\%)  was calculated, according to \cite{Helene}, as
  the  value for which: 

\begin{equation}
\frac{\displaystyle 2}{\sqrt{\pi}}
\int_{n_{\mu}(95\%)}^{\infty}
\frac{ e^{-(n_{\mu}-\overline{n_{\mu}})^2/2{\sigma}^2}}{\sqrt{2}{\sigma}}
\mathop{\mathrm dn_{\mu} = 0.05} \\ \\
\end{equation}

with $\overline{n_{\mu}}$=$n_{obs}-n_{exp}$ and ${\sigma}^2$=$n_{exp}$.

\item[-] 

 $A_{eff}(i)$ is the average effective detector area for every bin.
 It was computed by averaging the projected area seen by each muon and
 taking into account the geometrical and the tracking reconstruction
 efficiencies:

\begin{equation}
\mathop{\mathrm A_{eff}}(i)=
\frac{\displaystyle 1}{n_{obs}(i)}
\sum_{j=1}^{n_{obs}(i)}A({RA}_{i},{\delta}_{i})   \\ \\
\end{equation}

where ${RA}_i$ and ${\delta}_i$ are the muon arrival right ascension
and declination. 

\item[-] $T_{exp}$ is the exposure time computed as the time in which
  each  bin is visible. A bin is not  visible in our apparatus,
  if its zenith angle is larger than 72\degr.

\item[-]K is a correction factor which takes into account the fraction
  of muons within the bin dimensions for an hypothetical source placed
  at  the bin center. We calculated K=0.5 for a search cone of
  0.5\degr   ~half-angle and K=0.72 for 1.0\degr ~half-angle.

\end{enumerate}

The distribution of the 95\% C.L. upper limit $J_{\mu}^{stdy}$(95\%)
for all the  37176 bins in the sky region accessible for MACRO has an
average value of $2.3\times 10^{-13}\,\hbox{${\rm cm^{-2}s^{-1}}$}$. For almost all
bins,  the upper limit ranges between 
$1\times 10^{-13} \le J_{\mu}^{stdy} (95\%) \le 4 \times 10^{-13} \,\hbox{${\rm cm^{-2}s^{-1}}$} $.

Finally we also investigated some selected interesting point
sources (in Table \ref{tab:Nik}) identified by surface
telescopes and EAS  arrays \cite{bo5,bo6,bo7,EASTOP}. None of the 
  selected sources  in the list exhibits a significant deviation from
the  background.

\section{Searches for modulated signals from CygX-3 and HerX-1}

Periodic UHE gamma ray sources are attractive from the observational
point of view.  When  it is known that the emission from the source is
modulated  with a certain periodicity, the signal-to-noise ratio is
improved  by $\sqrt{N}$ \cite{bo8} by folding the event arrival
time modulo  the source period into N bins. This folding procedure was
employed  to analyze the data from the direction of CygX-3 and
HerX-1  \cite{Muraki,bo11,bo12}.

\begin{figure}
\begin{center}
\resizebox{15cm}{15cm}{\includegraphics{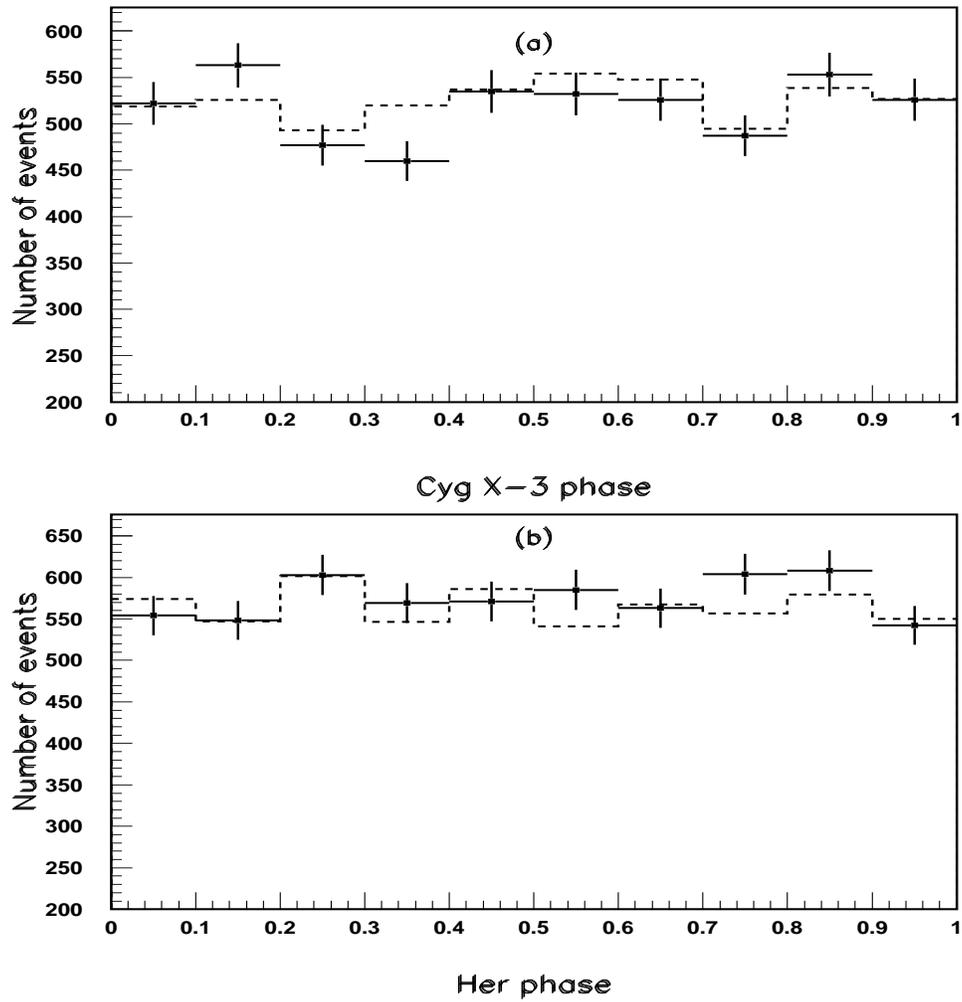}}
\end{center}
\caption{
Phase diagrams for muon events from 1\degr ~half-angle cone 
centered on (a) Cyg X-3 and (b) Her X-1. The dashed histograms
represent the simulations  of the background.
}
\label{fig:Hassane1}
\end{figure}

We used the quadratic ephemeris reported in Table~\ref{tab:hassane1}
and obtained
  from the fit to the observed  X-ray light curve \cite{bo9}, \cite{bo9a}.
The phase diagrams for events coming from a cone of 1\degr,
centered on the  position of the two sources, are shown in
 Fig.~\ref{fig:Hassane1}.
The expected number of background muons in each phase bin is also
  shown with the  dashed lines. 

\begin{figure}
\vspace{1cm}
\resizebox{15cm}{15cm}{\includegraphics{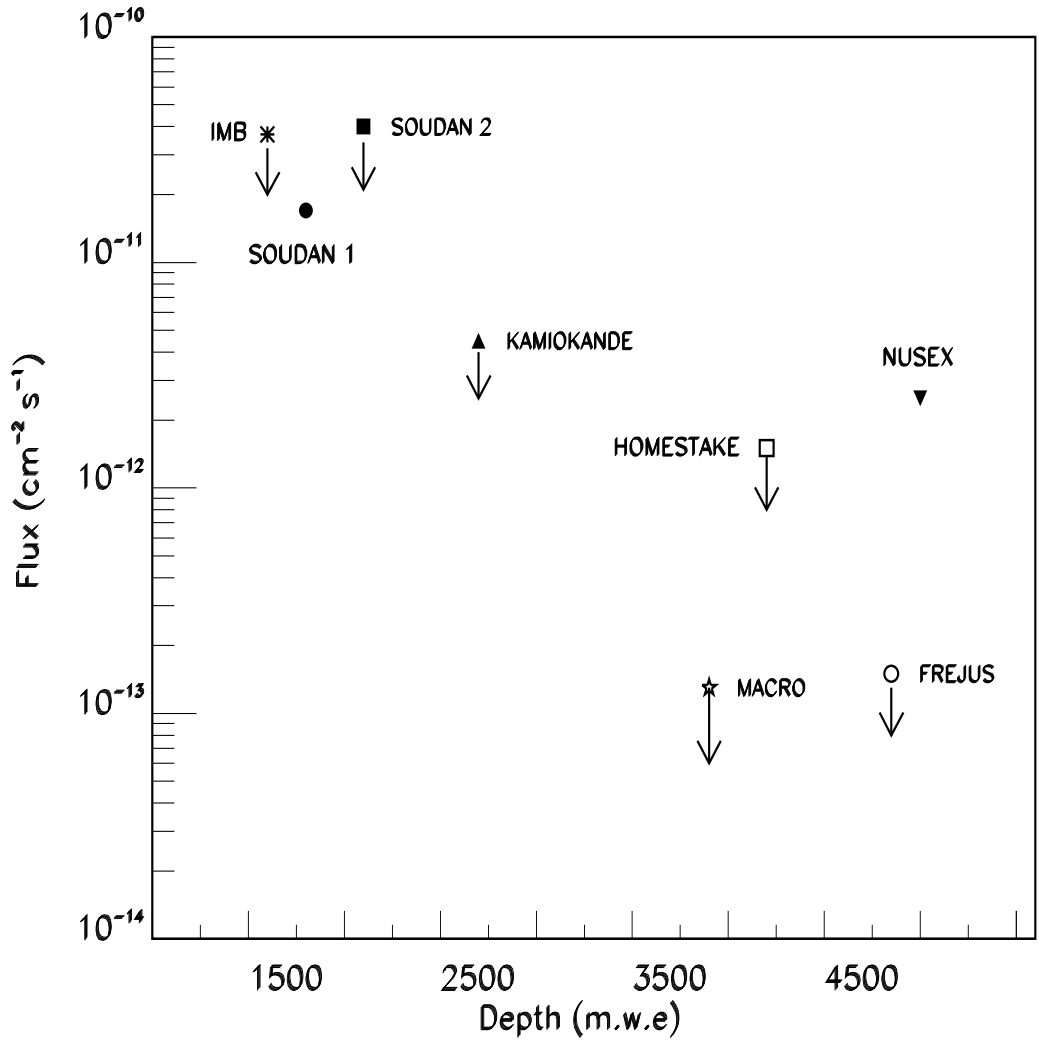}}
\caption{Searches for modulated signals from Cyg X-3: present situation for
the flux limits reported by other experiments
\cite{Soudan1,NUSEX,kamiok,frejus,imb,homestake} at 95\% C.L.  For MACRO, the 
method from ref.\cite{Helene} was used.
}
\label{fig:Hassane2}
\end{figure}

 The largest positive deviation ({1.8} $\sigma$) in the CygX-3 cycle
is  in the phase bin 0.1-0.2. The largest positive deviation for
HerX-1  (1.7 $\sigma$) is in the phase bin 0.7-0.8. Using the values  of the 
largest fluctuation we computed the 95$\%$
C.L. upper limits  to the modulated muon flux using equation
\ref{eq:helene},
 and the results are reported in Table~\ref{tab:hassane1}. 
In Fig.~\ref{fig:Hassane2} our computed upper limit for the
modulated emission  from CygX-3 is compared with those of previous
 underground experiments.
Our level is the lowest reached by such detectors.
The problem of understanding past positive observations still remain unsolved.

\section{Search for flaring activity from Mkn421 and Mkn501}

During the 1995-1999, several \v{C}erenkov telescopes reported
observations  of flaring activities up to 10-20 TeV from the two
celestial  objects Mkn421 and Mkn501\cite{Quinn2,Bradbury,Atai,Amenomori}.
These observations prompted us to search for bursting muon
signals in a  1\degr ~half-cone, around the position of these sources
of  U.H.E. photons. We  used two different methods.

\begin{figure}
\vspace{1cm}
\resizebox{15cm}{15cm}{\includegraphics{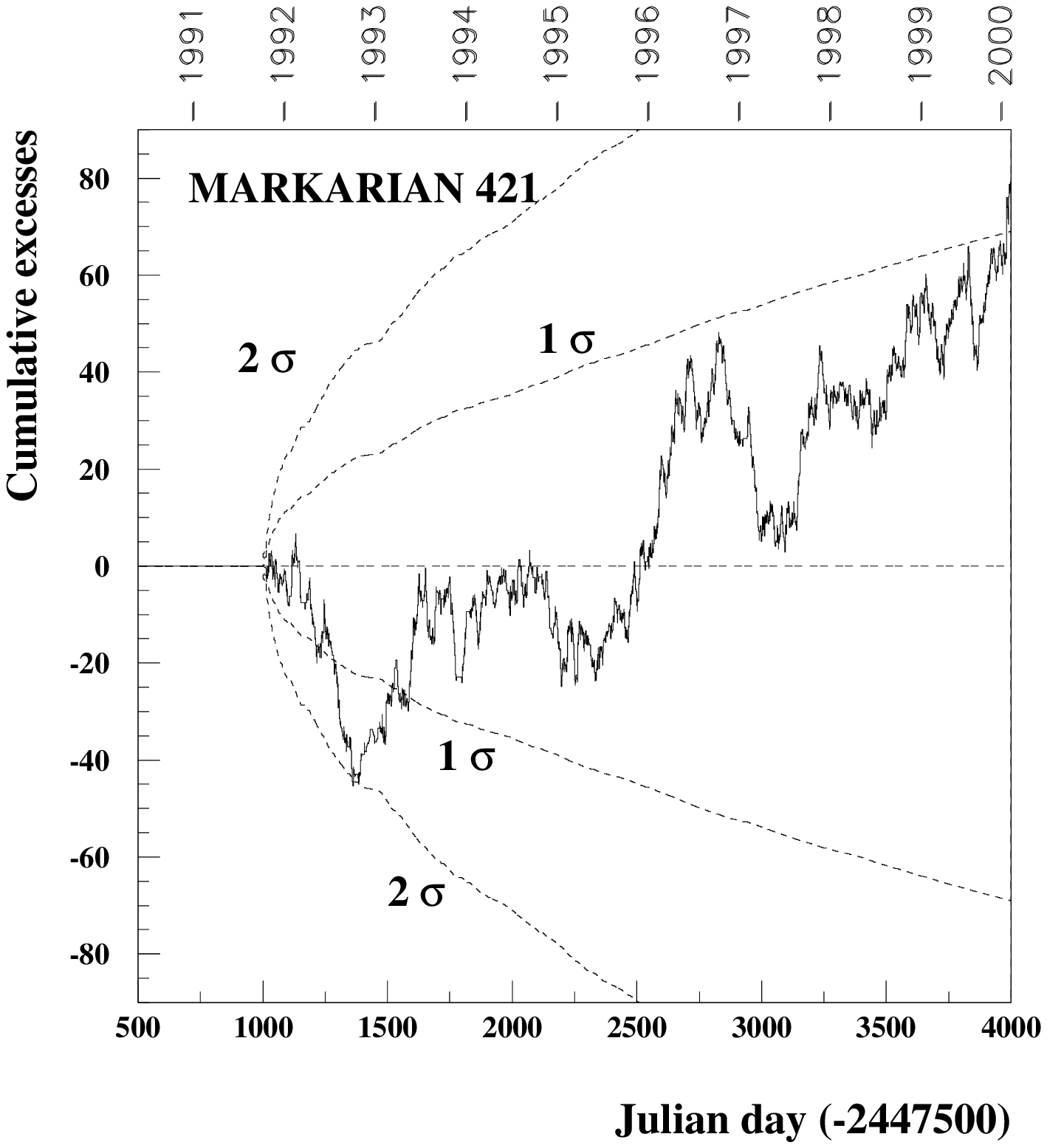}}

\caption{Cumulative muon excesses from the direction of Mkn421 (1\degr
   half-angle).}
\label{fig:burst}
\end{figure}

\begin{figure}
\vspace{1cm}
\resizebox{15cm}{15cm}{\includegraphics{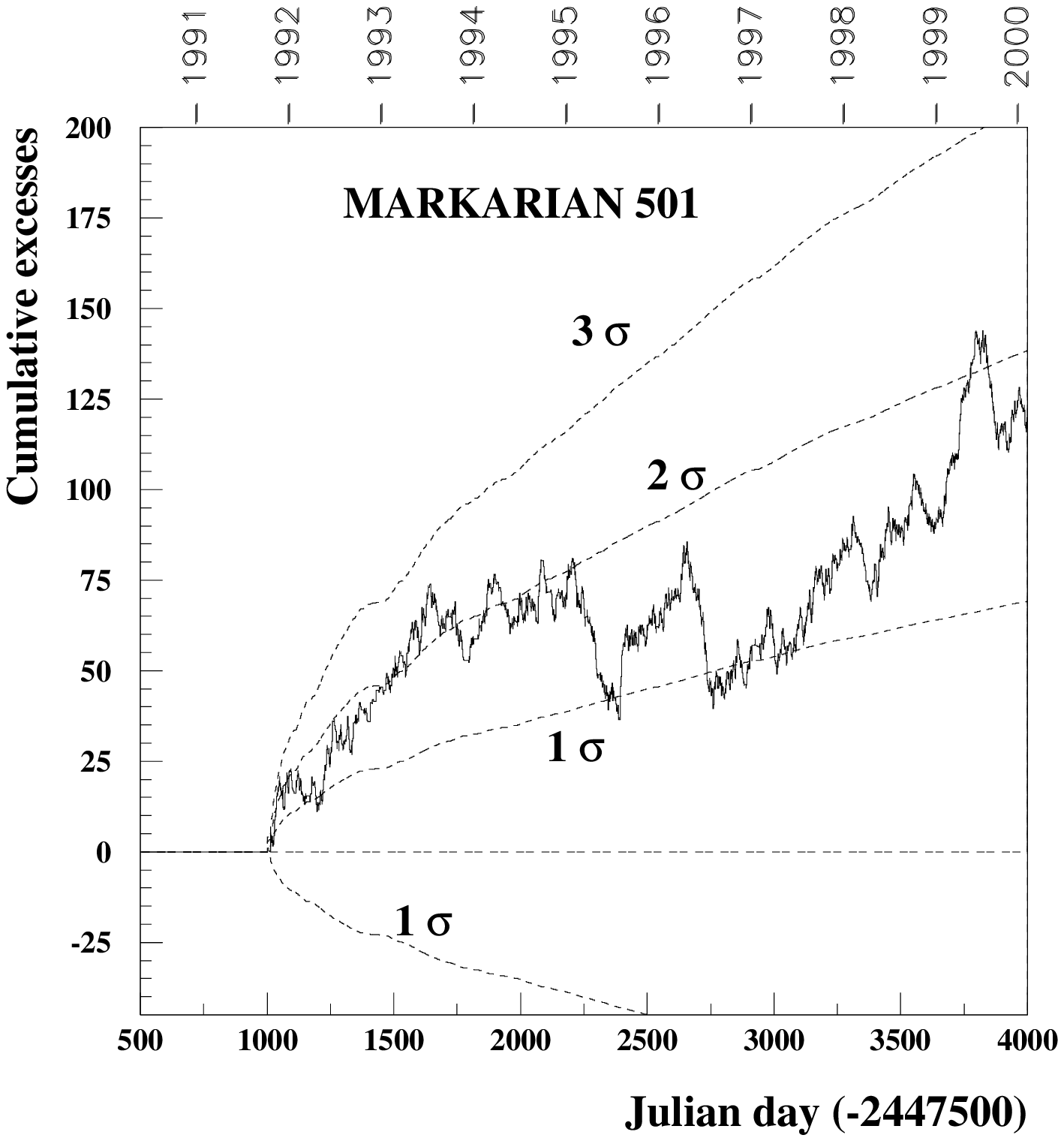}}
\caption{Cumulative muon excesses from the direction of Mkn501 ($1^{\circ}$
~half-angle).}
\label{fig:burst1}
\end{figure}

In the first method we studied, similarly to \cite{centA}, the
accumulation  rate of events coming from each of  the two
sources,  adding day-by-day the differences between the 
  measured  number of events and the calculated background.
Figures~\ref{fig:burst} and ~\ref{fig:burst1} show the cumulative
excesses as  a function of date (Julian days) since MACRO starting data
taking.  The cumulative excess presents fluctuations, but 
  never becomes  significant. 

In the second method we assumed (as in ~\cite{hegra}) that the
background  has a Poissonian distribution. If $n_{exp}$ is the
expected daily  background from the direction of a source, then the
probability to observe a random fluctuation of the background large as
the observed $n_{obs}$ events in a day, is given by:
\begin{equation}
\mathop{{P}= 1- 
\sum_{n=0}^{n_{obs}-1}   
\frac {\displaystyle \alpha^n}{\displaystyle (1+\alpha)^{n+n_{exp}+1}\qquad}
\frac {\displaystyle (n+n_{exp})!}{\displaystyle n!\times n_{exp}!\qquad}}
\end{equation}

\noindent where $\alpha$ is  the ratio of the ON-source time to the
OFF-source  time. Because we extract 25 simulated background events
for  each real one,  $\alpha$ is 0.04. 
For a Poissonian background, the cumulative frequency distribution of P
is  expected to be a power law with index -1, in logarithmic scale.

\begin{figure}
\vspace{1cm}
\resizebox{15cm}{15cm}{\includegraphics{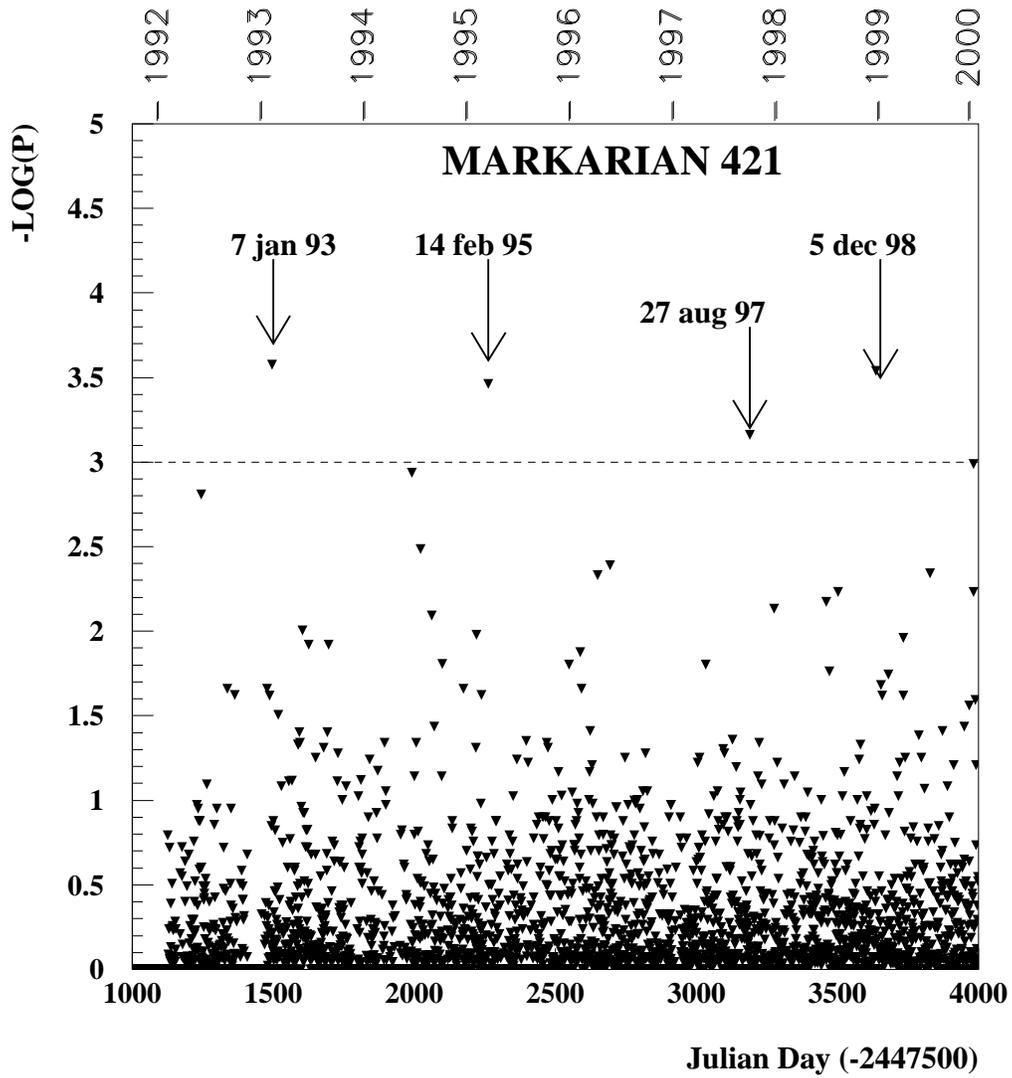}}
\caption{
Value of the quantity $-\log P$ (defined in the text) evaluated
day-by-day from the direction  of Mkn421 using a search cone with
1\degr half-angle.  
The days with a value of $-\log P$ exceeding the
probability value of $10^{-3}$ that we have choosen as the  
attention level  (indicated with the dashed line) are indicated by the arrows.
}
\label{fig:lam1}
\end{figure}

\begin{figure}[p]
\vspace{1cm}
\resizebox{15cm}{15cm}{\includegraphics{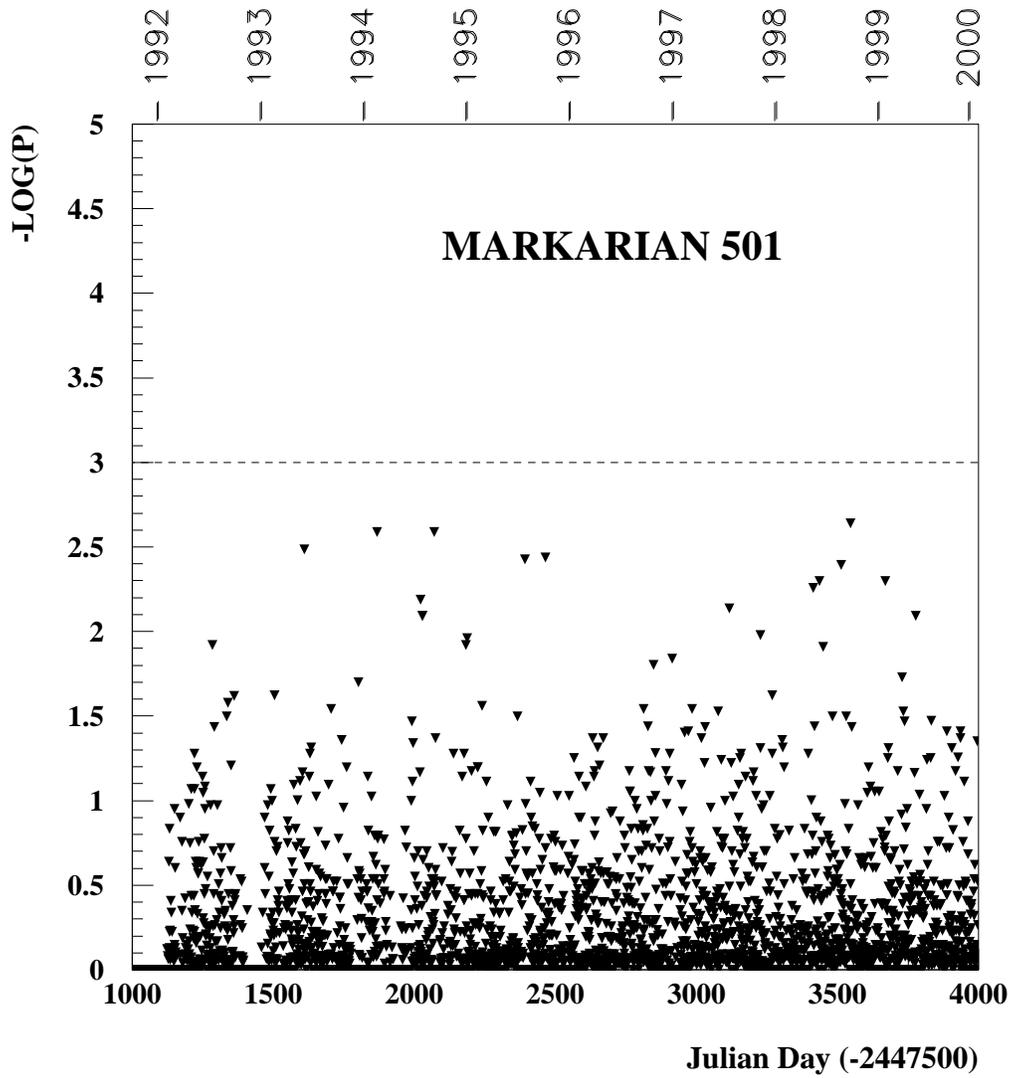}}

\caption{
The same of the previous figure, but for the Mkn501 source.  No days
exceeded  the
probability value of $10^{-3}$ that we have choosen as  the  attention level.
}
\label{fig:lam2}
\end{figure}

Figures~\ref{fig:lam1} and ~\ref{fig:lam2} show the value of -$\log P$
evaluated day-by-day for Mkn421 and Mkn501, respectively. 
In Fig.~\ref{fig:lam1} the date of the largest fluctuations for Mkn421
with  respect to the background are also indicated.

\begin{figure}
\vspace{1cm}
\resizebox{15cm}{15cm}{\includegraphics{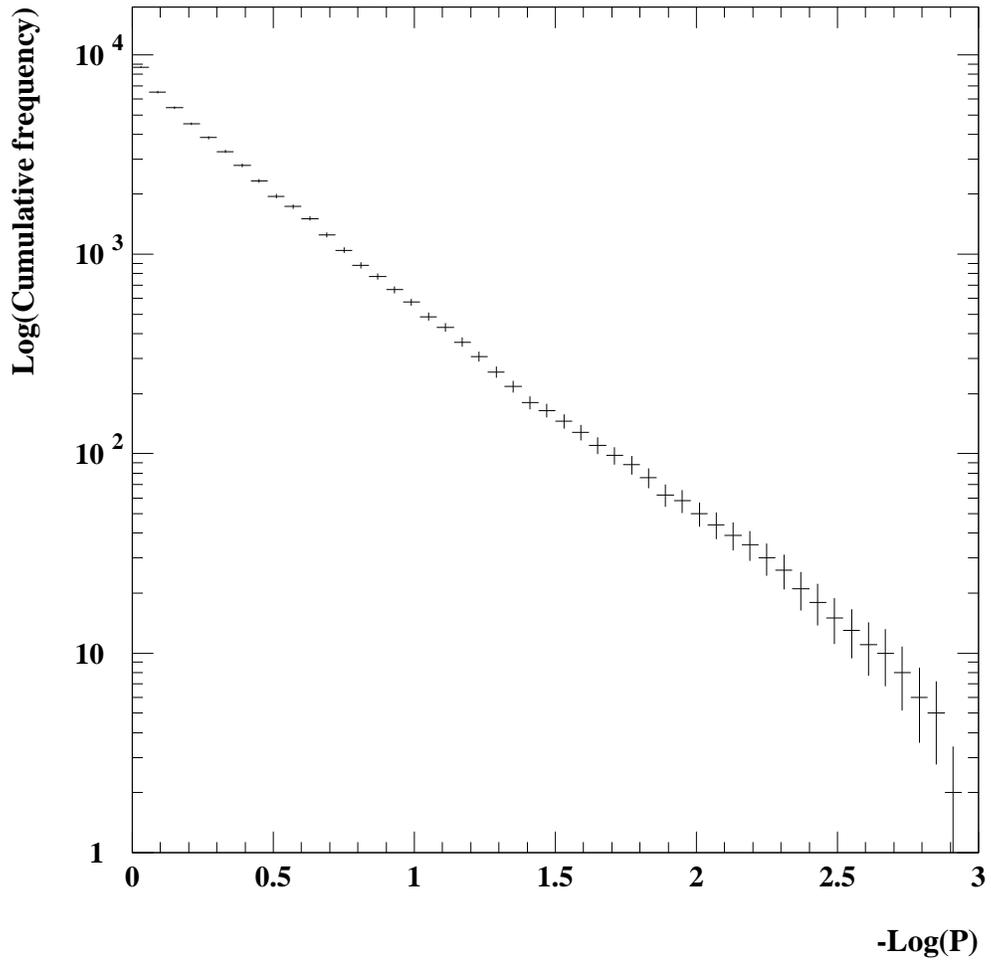}}
\caption{
The cumulative distribution of the $-\log P$ quantity (see text) for all  
bins of the all-sky survey.
}
\label{fig:skylam}
\end{figure}

To verify
the probability of such a positive fluctuation from
Mkn421, we computed the quantity -$\log P$ for a set of 
selected bins having an exposure similar to the two
Mkn objects, each monitored for about 3600 days.
 Figure~\ref{fig:skylam} shows the log-log plot of
the  cumulative frequency distribution for all these sky bins.  
Since this
cumulative distribution has a slope close to -1, as expected in the
case of  no source detected, we can use the Poissonian statistics 
to compute the expectation to see large fluctuations.

We set as an {\it attention level} a probability of $\le 10^{-3}$ for
a fluctuation.
We found 4 days with probability larger than $10^{-3}$ for Mkn421,
and none  for Mkn501. We observed the Mkn421 source for 3600 days:
assuming  as attention level for any source the probability value of
$10^{-3}$, the expected number of random fluctuations with probability
lower  than $10^{-3}$ is $3600 \cdot (1\times10^{-3})=3.6$. Therefore the Poissonian probability to observe
4 random  fluctuation {in the same period}, with an average of
3.6,  is about 20$\%$. This probability value therefore excluded a
 positive observation of a burst from this source.

\section{Conclusions}

Since February 1989, the MACRO detector collected 49.9 million 
well-constructed  muons. Using this sample, we searched for muon
excesses  above background from all visible sky directions and from
known  astrophysical sources.
No significant excesses were found from the all-sky survey.
We computed the 95\% confidence level upper limit $
J^{stdy}_{\mu}(95\%)$  for a steady muon flux for all the 
37176 sky cells; the average  value of $ J^{stdy}_{\mu}(95\%)$ is equal to
2.3$\times$$10^{-13}$  ${\rm cm^{-2}~s^{-1}}$. We analyzed the muons coming
from  the direction of Cyg X-3 and Her X-1, searching for a modulated
emission, with a negative result. 
The search for a steady or episodic emission coming from
Mkn421  and Mkn501 was
made  with two different methods. We found no muon excess above the
estimated  background.

\section{Acknowledgments}
We gratefully acknowledge the support of the director and of the staff of
the Laboratori Nazionali del Gran Sasso and the invaluable assistance of the
technical staff of the Institutions participating in the experiment. We
thank the Istituto Nazionale di Fisica Nucleare (INFN), the U.S. Department
of Energy and the U.S.  National Science Foundation for their generous
support of the MACRO experiment. We thank INFN, ICTP (Trieste), WorldLab and
NATO for providing fellowships and grants (FAI) for non Italian citizens.

This work is dedicated to the memory of Lynn Miller.

\newpage
\begin{table}[h]
\begin{center}
\begin{tabular}{||c|c|r|r||} \hline
Period & Number of & Number of &Exposure \\ 
&  SuperModules & events & Time (hr)\\ \hline
\small Feb,27 1989-May,20 1989 & 1     &244333 & 1942.9 \\
\small Nov,11 1989-May,10 1990 & 1     &365148 & 3072.1\\
\small May,10 1990-Jul,5 1991  & 1,2   &1308311 &5274.9 \\
\small Jul,5 1991-Apr,29 1994  & 1-6   &11549606 &16247.8 \\
\small Apr,29 1994-Dec,15 2000 & 1-6+A &36490390 &47535.4 \\ \hline
\end{tabular}
\end{center}
\caption{Description of the muon data sets collected by  MACRO.
As construction of MACRO progressed, more supermodules came on-line
and increased MACRO's collection area. Most of MACRO's muon data was
collected after MACRO was fully operational in 1994.
}
\label{tab:sample}
\end{table}

\newpage
\begin{table}[h]
\begin{tabular}{|r|c|c|c|c|c|c|}
\hline
        &           &         &         & Area  &$T_{exp}$ &  Flux         \\
        & $n_{obs}$ &$n_{exp}$&$\sigma$ &       & $10^6$   &   $10^{-13}$          
\\
        &           &         &         &$m^2$  & sec      &$cm^{-2}s^{-1}$\\
\hline\hline
MKN421 (0.5\degr)  &  1373     &  1382.8   &  -0.26    & 745   &  162     & 1.6 
\\
        (1.0\degr) &  6506     &  6389.7   &  1.5    & 745   &  162     & 3.7\\
\hline
MKN501 (0.5\degr)  &  1482     &  1429.9 &  1.4    & 739   &  164     & 2.4 \\
 (1.0\degr)        &  6544     &  6459.5 &  1.1    & 739   &  164     & 3.4 \\
\hline
CRAB  (0.5\degr)   &  1685     &  1651.5 &  0.8    & 744   &  137     & 2.7\\
 (1.0\degr)        &  6808     &  6764.0 &  0.5    & 744   &  137     & 3.6\\
\hline
Cyg X-3 (0.5\degr) &  1376     &  1382.8 & -0.17   & 740   &  164    & 1.6\\
       (1.0\degr)  &  6340     &  6370.7 & -0.4    & 740   &  164    & 2.4\\
\hline
Her X-1 (0.5\degr) &  1521     &  1501.7 &  0.5    & 733   &  159     & 2.1\\
        (1.0\degr) &  6530     &  6519.0 &  0.14    & 733   &  159    & 2.8\\
\hline
3C66A (0.5\degr)   &  1378     &  1354.1 &  0.65     & 737   &  168    & 1.9 \\
      (1.0\degr)   &  6398     &  6305.3 &  1.17    & 737   &  168    & 3.4\\
\hline
1ES514 (0.5\degr)  &  1550     &  1618.6 &  -1.7    & 729   &  180     & 1.2\\
       (1.0\degr)  &  6969     &  7003.8 &  -0.4    & 729   &  180     & 2.3\\
\hline
QX SGE  (0.5\degr) &  1731     &  1719.2 &  0.28    & 770   &  137    &2.3\\
        (1.0\degr) &  7170     &  7083.1 &  1.03    & 770   &  137 &3.9 \\     
\hline
SS Cyg  (0.5\degr) &  1382     &  1406.5 & -0.65    & 735   &  169  &1.4 \\
        (1.0\degr) &  5548     &  5526.7 &  0.28    & 735   &  169  &2.5 \\
\hline
Geminga (0.5\degr) &  1709     &  1692.8 &  0.39    & 748   &  125  &2.7\\
         (1.0\degr)&  7632     &  7693.3 &  -0.7    & 748   &  125  &3.1 \\
\hline
3C273   (0.5\degr) &  1138     &  1168.1 & -0.88    & 636   &  96  &2.5\\ 
         (1.0\degr)&  4529     &  4625.7 & -1.4     & 636   &  96  &3.1 \\
\hline
3C279    (0.5\degr)&  645      &  657.2  & -0.48    & 569   &  90  & 2.5\\
         (1.0\degr)&  2644     &  2706.3 & -1.2     & 569   &  90  & 2.9\\
\hline
2CG095   (0.5\degr)& 1709      &  1724.9 & -0.38    & 729   &  180 & 1.6\\
           (1.0\degr)& 6919    &  7003.1 & -1.0     & 729   &  180 & 1.9 \\
\hline
2CG135   (0.5\degr)& 1892      &  1863.8 & 0.65     & 712   &  154 &2.6\\ 
            (1.0\degr)& 7946   &  7899.5 & 0.52     & 712   &  154 &3.5 \\
\hline
4U1907  (0.5\degr) & 1527      &  1543.2 & -0.4     & 698   &  99  & 2.8\\
         (1.0\degr)&  6566     &  6496.8 &  0.86    & 698   &  99 &5.5 \\
\hline
4U0115  (0.5\degr) & 2024      &  2033.7 & -0.21    & 711   &  145 & 2.3\\
         (1.0\degr)& 8351      &  8489.5 & -1.5     & 711   &  145 & 2.4\\
\hline
V1341   (0.5\degr) &1382       &  1406.5 & -0.65    & 748   &  161&1.5 \\
         (1.0\degr)&5481       &  5560.9 & -1.1     & 748   &  161 &1.9 \\
\hline
PSR1929  (0.5\degr)& 1559      & 1578.7  & -0.5     & 716   &  106 &2.6\\ 
         (1.0\degr)& 6012      & 5981.7  & 0.4      & 716   &  106 &4.3 \\
\hline
PSR1855  (0.5\degr)&  1567     &  1580.8 & -0.34    & 716   & 107&2.6\\ 
          (1.0\degr)& 6576     &  6517.6 &  0.7     &716   & 107 &4.8 \\ 
\hline          
\end{tabular}
\caption{Search for muon excesses from selected sources, using
  half-cones of $0.5^\circ$ and $1^\circ$. The number of muons
  observed $n_{obs}$ and expected $n_{exp}$ are tabulated, with the
  quantity $\sigma=\frac{(n_{obs}-n_{exp})}{\sqrt{n_{exp}}}$ to
  indicate the significance of deviations from expected. The area,
  exposure time and calculated upper limit on the muon flux (95\%
  ~C.L.) is also shown.} 

\label{tab:Nik}
\end{table}

\newpage
\begin{table}[h]
\small    
\begin{center}
\begin{tabular}{|c|c|c|c|} 
\hline
\multicolumn{1}{|c}{} &    &        Cyg X-3   \cite{bo9}  &Her X-1 \cite{Deeter} 
\\ \hline \hline
    &P [d] & 0.19968271 & 1.700167788   \\ 
    &      &  ($\pm$2.4$\times10^{-7}$) & ($\pm$1.1$\times10^{-8}$)  \\
\cline{2-4}
Ephemeris  &$\dot{P}/P$  [d/yr] & $(1.17 \pm 0.44) \times 10^{-6}$   & 
                                                         $< 2 \times 10^{-8}$ \\
\cline{2-4}
 & $T_0$ [JD]  &   2440949.8989 & 2442859.726688 \\
 &     &   ($\pm$0.0012) & ($\pm$7$\times 10^{-6})$ \\
\hline\hline
Flux  &$J_\mu^{mod}(95\%)$ &$1.4 \times 10^{-13}$&$1.6 
\times 10^{-13}$ \\
limits 1\degr   &(${\rm cm^{-2}s^{-1}}$) & &\\ 
\hline
\end{tabular}
\end{center} 
\caption{
 Ephemeris parameters for Cyg X-3 and Her X-1 and modulated
      flux limits computed using equation \ref{eq:helene}   in a 1$^\circ$
      cone   around the source position.}
\label{tab:hassane1}
\end{table} 




\clearpage

\end{document}